\documentclass[
    reprint,
    superscriptaddress,
    amsmath,
    amssymb,
    aps,
    prb,
    floatfix,
]{revtex4-2}

\usepackage[T1]{fontenc}
\usepackage{babel}
\usepackage{graphicx}
\usepackage{dcolumn}
\usepackage{bm}
\usepackage{svg}
\usepackage{import}
\usepackage{xcolor}
\usepackage[normalem]{ulem}
\usepackage{hhline}
\usepackage{amsmath}
\usepackage{hyperref}
\hypersetup{
    colorlinks=true,
    linkcolor=blue,
    filecolor=magenta,      
    urlcolor=blue,
    citecolor=blue,
}
\usepackage{siunitx}
\DeclareSIUnit \belm {Bm}
\usepackage{braket}
\usepackage{nicefrac}
\usepackage{placeins}

\begin{document}

\preprint{APS/123-QED}

\title{Miniaturized magnetic-field sensor based on nitrogen-vacancy centers}

\author{Stefan Johansson}
\thanks{Both authors contributed equally}
\affiliation{Department of Physics and State Research Center OPTIMAS, RPTU University Kaiserslautern-Landau, Erwin-Schroedinger-Str. 46, 67663 Kaiserslautern, Germany}
\author{Dennis Lönard}
\thanks{Both authors contributed equally}
\affiliation{Department of Physics and State Research Center OPTIMAS, RPTU University Kaiserslautern-Landau, Erwin-Schroedinger-Str. 46, 67663 Kaiserslautern, Germany}
\author{Isabel Cardoso Barbosa}
\affiliation{Department of Physics and State Research Center OPTIMAS, RPTU University Kaiserslautern-Landau, Erwin-Schroedinger-Str. 46, 67663 Kaiserslautern, Germany}
\author{Jonas Gutsche}
\affiliation{Department of Physics and State Research Center OPTIMAS, RPTU University Kaiserslautern-Landau, Erwin-Schroedinger-Str. 46, 67663 Kaiserslautern, Germany}
\author{Jonas Witzenrath}
\affiliation{Department of Physics and State Research Center OPTIMAS, RPTU University Kaiserslautern-Landau, Erwin-Schroedinger-Str. 46, 67663 Kaiserslautern, Germany}
\author{Artur Widera}
\email{widera@physik.uni-kl.de}
\affiliation{Department of Physics and State Research Center OPTIMAS, RPTU University Kaiserslautern-Landau, Erwin-Schroedinger-Str. 46, 67663 Kaiserslautern, Germany}

\date{\today}

\begin{abstract}
\noindent The nitrogen-vacancy (NV) center in diamond is a prime candidate for quantum sensing technologies.
Here, we present a fully integrated and mechanically robust fiber-based endoscopic sensor with a tip diameter of $\SI{1.25}{\milli\meter}$.
On its tip, a direct laser writing process is used to fixate a diamond containing NV centers above the fiber’s core inside a polymer structure.
Additionally, a metallic direct laser-written antenna structure next to the fiber facet allows efficient microwave manipulation of NV center spins.
The sensor achieves a shot-noise-limited magnetic-field sensitivity of $\SI{5.9}{\nano\tesla \per \sqrt{\hertz}}$ using a $\SI{15}{\micro\meter}$-sized microdiamond at a microwave power of $\SI{50}{\milli\watt}$ and optical power of $\SI{2.15}{\milli\watt}$.
Using lock-in techniques, we measure a sensitivity of $\SI{51.8}{\nano\tesla \per \sqrt{\hertz}}$.
Furthermore, we introduce a dual-fiber concept that enables, in combination with a direct laser-written structure, independent guiding of excitation and fluorescence light and thus reduces background autofluorescence.
Moreover, controlled guiding of excitation light to the diamond while avoiding sample illumination may enable operation in light-sensitive environments such as biological tissue.
While the demonstrated sensitivity is achieved using a single-fiber configuration, the dual-fiber approach provides a path towards integrating smaller diamonds, where autofluorescence would otherwise limit performance.
We demonstrate the capability of vector magnetic field measurements in a magnetic field as used in state-of-the-art ultracold quantum gas experiments, opening a potential field in which high resolution and high sensitivity are necessary.

\end{abstract}

\maketitle

\section{Introduction}
The negatively charged nitrogen-vacancy (NV) center in diamond is an atomic-scale point defect in the diamond lattice consisting of a vacancy and a substitutional nitrogen atom next to it \cite{loubser_electron_1978, doherty_nitrogen-vacancy_2013}. 
It provides optically addressable spin states and allows optical readout of the spin-state population \cite{rondin_magnetometry_2014}. 
When resonantly excited with a microwave (MW) field, it enables, due to spin-dependent intersystem crossing, optically detected magnetic resonance (ODMR) spectroscopy \cite{levine_principles_2019}.
The inclusion of the defect in the diamond lattice results in a solid-state sensor with atom-like behavior \cite{haque_overview_2017}, making it a leading candidate in research as a highly sensitive vectorial magnetic-field sensor with high spatial resolution \cite{barry_sensitive_2024} and the capability to measure temperature \cite{acosta_temperature_2010, choe_precise_2018}, pressure \cite{kehayias_imaging_2019}, and electrical fields \cite{dolde_electric-field_2011, van_oort_electric-field-induced_1990, tamarat_stark_2006, tetienne_quantum_2017}.

Other magnetic-field sensors such as superconducting quantum interference devices (SQUIDs) and optically pumped atomic magnetometers (OPMs) have demonstrated sensitivities in the $\SI{}{\femto\tesla \per \sqrt{\hertz}}$ range \cite{storm_ultra-sensitive_2017, griffith_femtotesla_2010}.
However, they often require cryogenic temperatures or are not intrinsically capable of measuring vectorial magnetic fields with a single sensing element.
Many applications, like measurements of electrical currents in coils or high-power electronics, paramagnetic materials, or permanent magnetic materials, need to be measured accurately with $\SI{}{\nano\tesla\per\sqrt{\hertz}}$ sensitivities \cite{scholten_widefield_2021, hatano_high-precision_2022, broadway_imaging_2020, steinert_magnetic_2013}.
Often, they require only moderate absolute sensitivities but strongly profit from an increased sensitivity per distance \cite{degen_microscopy_2008, rembold_introduction_2020, schafer-nolte_eiker_oliver_development_2014}.
Diamonds containing NV centers provide high overall sensitivities, especially when combined with the use of flux concentrators \cite{fescenko_diamond_2020}, $^{12}\mathrm{C}$-enriched diamonds for longer coherence times \cite{barry_sensitive_2024} or by a large number of spins \cite{wolf_subpicotesla_2015, shim_multiplexed_2022}, and are usable from cryogenic temperatures up to hundreds of $\SI{}{\celsius}$ \cite{kubota_wide_2023, scheidegger_scanning_2022, liu_coherent_2019}, promising to be the next generation of magnetic-field sensors.
Moreover, single diamond-based sensors have demonstrated vector magnetometry with a wide dynamic range and high sensitivity \cite{scholten_widefield_2021}.
Current state-of-the-art approaches to measuring magnetic fields are based on stationary setups such as wide-field microscopy utilizing diamond plates \cite{abrahams_integrated_2021}, scanning diamond probes \cite{appel_fabrication_2016}, or integrated portable devices \cite{webb_nanotesla_2019}.
While stationary scanning probe and wide-field setups often provide very high magnetic-field sensitivities or high spatial resolution, these lab setups can not be used as practical, portable sensors due to their size.
Furthermore, most scanning probe setups do not provide a vectorial field due to a single NV center, and, just like confocal setups, stable environmental conditions are required for reliable results.

Thus, we focus on portable, especially fiber-based, approaches for NV-sensors in this work.
The reported portable devices can be split into two groups.
Either they are fully integrated, consisting of necessary excitation and detection elements such as MW antennas, optical excitation light sources, and detectors close by, or directly attached \cite{zheng_hand-held_2020, sturner_compact_2019, pogorzelski_compact_2024}.
Integrating these elements, such as detectors and excitation sources, limits the possible miniaturization of the actual sensor.
Alternatively, they consist of two parts where the sensor's head, as well as the excitation and detection sections, are separated with optical fibers and electrical wires \cite{patel_subnanotesla_2020, kuwahata_magnetometer_2020, sturner_integrated_2021, kim_cmos-integrated_2019, zhao_all_2023}.
Therefore, miniaturization can be enhanced by only integrating the necessary elements on a sensor's tip and connecting them to wires and optical fibers. 
However, combining an NV-doped diamond with optical and microwave access on a compact form factor leads to challenges in finding a robust platform.
Usually, fiber-based sensors use light-curable adhesive to attach diamonds to a fiber, often with mismatching sizes between the diamond and optical fiber, sacrificing optical performance due to non-ideal light-guiding efficiencies.
Often, the integration of an antenna is realized by attaching an external antenna or a piece of wire using tape or glue, leading to a fragile sensor and poor MW efficiency \cite{fedotov_fiber-optic_2014, zhao_efficient_2023, homrighausen_microscale_2024}.
Other publications use randomly oriented diamond ensembles, which only provide scalar field information and often have worse resolution than single diamond sensors or lack the integration of a microwave-emitting structure \cite{dong_fiber_2018, filipkowski_magnetically_2022, chen_nanodiamond-based_2022}.
Previously reported results also suggest that positioning the microwave-emitting element near the diamond can provide a good sensitivity even at low power settings \cite{dix_fiber-tip_2022}.

In this context, a further limitation of miniaturized optical fiber-based sensors is the so-called autofluorescence of optical fibers \cite{bianco_comparative_2021, udovich_spectral_2008, yufanli_2023}, especially when the diamond size is reduced below a few hundred nanometers. 
This autofluorescence arises from the necessary excitation light required for NV centers, which also excites photo-active compounds within the fiber and typically overlays spectrally with the fluorescence of the NV center.
When the diamond size and the number of NV centers are decreased, the total fluorescence from the actual sensor decreases while the autofluorescence remains constant for the same laser power.
This ratio leads to a worse signal-to-noise (SNR) ratio and, thereby, sensitivity, mainly when small sensing diamonds or diamonds with low NV center dopings are utilized.
Although bleaching strategies to reduce the autofluorescence intensity temporarily are suggested, it can not be eliminated completely \cite{Bianco_21}.

Here, we show a robust, miniaturized dual-fiber-based platform and sensor using standardized components and state-of-the-art additive manufacturing techniques.
We use metallic direct laser writing of a silver resist to create an antenna structure next to a direct laser-written polymer structure.
The polymer structure is used to guide excitation and fluorescence light and to fixate a single deterministically positioned $\SI{15}{\micro\meter}$-sized diamond to the tip of an optical fiber on a platform diameter of $\SI{270}{\micro\meter}$.
For our magnetic-field sensor, we find a shot-noise (SN) limited sensitivity of $\SI{5.9}{\nano\tesla \per \sqrt{\hertz}}$ in the single-fiber configuration with an intrinsic spatial resolution (based on the diamond size of $\SI{15}{\micro\meter}$), at a microwave power as low as $\SI{50}{\milli\watt}$ and optical laser power of $\SI{2.15}{\milli\watt}$ with a total volume of our sensor head of $\SI{52.8}{\cubic\milli\meter}$. 
For better comparison across sensors with diamonds of different sizes, we calculate a volume normalized (volumetric) shot-noise-limited sensitivity related to the diamond size of $\SI{7.8}{\pico\tesla \per \sqrt{\hertz} \times {\milli\meter}^{3/2}}$, assuming the microdiamond as a homogeneous $\SI{15}{\micro\meter}$-sized sphere.

As a proof of principle, our dual-fiber sensor approach paves the way towards the integration of smaller diamonds down to a size below $\SI{100}{\nano\meter}$ without being significantly affected by the autofluorescence in a detecting fiber.
This is achieved by using different optical fibers guiding excitation or NV-fluorescence light in combination with a connecting waveguide structure.
We measure a reduction of the amount of laser light guided in the detection fiber by \numrange{2}{3} orders of magnitude, resulting in a significant reduction of autofluorescence.
However, we note that a sensor utilizing a $\SI{15}{\micro\meter}$-sized diamond is not mainly limited by autofluorescence and imperfections in the polymer structure significantly impact the reached sensitivity.

In section \ref{sec:materials}, we first describe the steps to build the platform and the process to create the antenna structure and polymer waveguide, which also fixes the diamond. 
In Section \ref{sec:results}, measurements characterize our sensor regarding autofluorescence, sensitivity, and noise, and we show a practical application.
To better compare current sensor designs, we compare our approach to other fiber-based magnetic-field sensors in size and sensitivity and normalize measured sensitivities by the utilized diamond size.
Our results and limitations are summarized in section \ref{sec:conclusion}, and further challenges for integrating nanometer-sized diamonds are discussed.

\section{Materials and Methods}\label{sec:materials}
\begin{figure*}[htbp]
    \centering
    \includegraphics[width=\textwidth]{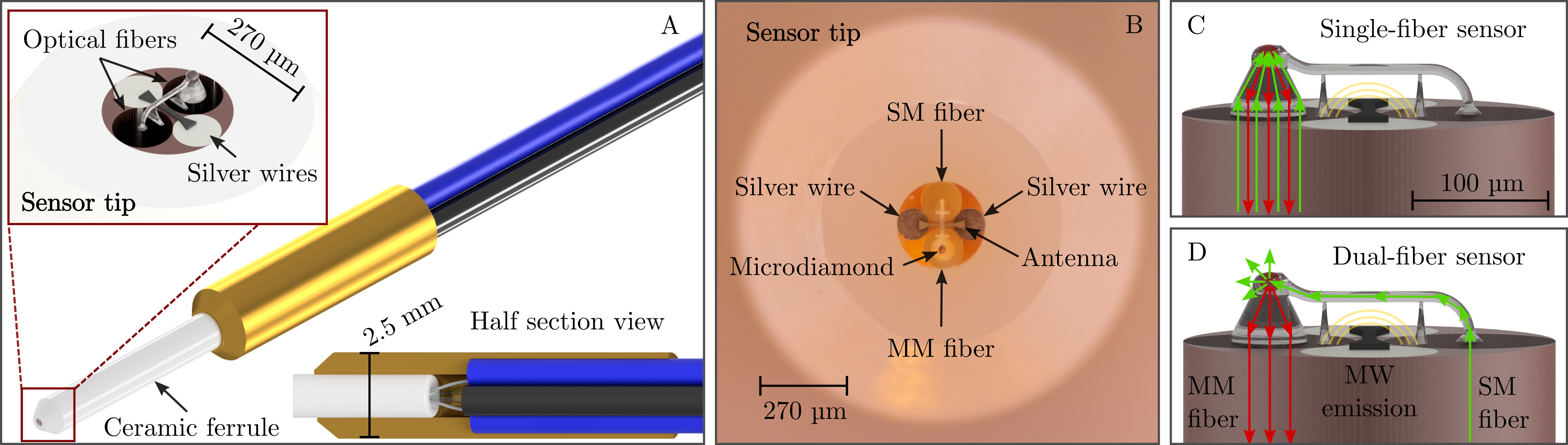}
    \caption{
        (A) Rendered images of the assembled sensor platform. Two silver wires and two optical fibers are inserted into a ceramic ferrule. As a reinforcement between the ferrule and the protective tubing of the fibers a mating sleeve is added. (B) Tip of a sensor after fabrication. (C) and (D) rendered images of the sensor tip with the antenna and a polymer waveguide structure and the light path for green excitation light and red fluorescence in a single-fiber and dual-fiber configuration.
    }
    \label{fig:assemblysensor}
\end{figure*}

Our sensor head is split into two components, the so-called sensor platform and the tip of the sensor.
The sensor platform connects the MW guiding wires and light-guiding optical fibers together in an optical fiber ferrule as depicted in FIG. \ref{fig:assemblysensor} (A). 
After filling the remaining space using epoxy and subsequent polishing of the tip, the small tip diameter enables the use of microscopic additive technologies such as direct laser writing at the tip of the fibers and wires, as shown in FIG. \ref{fig:assemblysensor} (B).

The platform also enhances the mechanical stability of the whole sensor without significantly increasing the size of the necessary parts of the sensor.
The silver antenna and polymer light guiding structure at the sensor's tip are created using a direct laser writing process based on a two-photon absorption process in a commercial lithography system (PPGT+, Nanoscribe).
We modified the system with a positioning stage, which allows inserting and positioning of a sensor tip on five axes above an NV-doped ($\approx\SI{3.5}{}\,\mathrm{ppm}$) $\SI{15}{\micro\meter}$-sized microdiamond (MDNV15umHi, Adamas Nanotechnologies) using the device's imaging system, as shown in FIG. \ref{fig:dlw}.
Using this platform, the sensor can be used as a single-fiber or dual-fiber sensor, as depicted in FIG. \ref{fig:assemblysensor} (C, D).
A detailed description of all steps of the fabrication is given in the following subsections.

\subsection{Sensor platform}\label{subsec:platform}
The sensor's fabrication starts with the platform, which consists of two optical fibers (FG050LGA and 460HP, Thorlabs) and two silver wires with a diameter of $\SI{100}{\micro\meter}$. 
Fibers and wires are then inserted into a ceramic ferrule with an inner diameter of $\SI{270}{\micro\meter}$ and an outer diameter of $\SI{1.25}{\milli\meter}$ (CFX270-10, Thorlabs), as depicted in FIG. \ref{fig:assemblysensor}(A). 
The fibers, wires, and ferrule are chosen to fit tightly, ensuring a repeatable arrangement, as shown in FIG. \ref{fig:assemblysensor}(A, B). 
Protective tubing for optical fibers (FT900KB and FT900KK, Thorlabs) jackets the silver wires and optical fibers.
The tubing is connected to the ceramic ferrule in a metal or polycarbonate sleeve, as shown in FIG. \ref{fig:assemblysensor}(A).
The sensor platform's elements are then fixed in position using epoxy (353NDPK, Thorlabs) and polished after curing using optical fiber polishing tools.

\subsection{Sensor tip fabrication}\label{subsec:sensortip}
\begin{figure}[htbp]
    \centering
    \includegraphics[width=\linewidth]{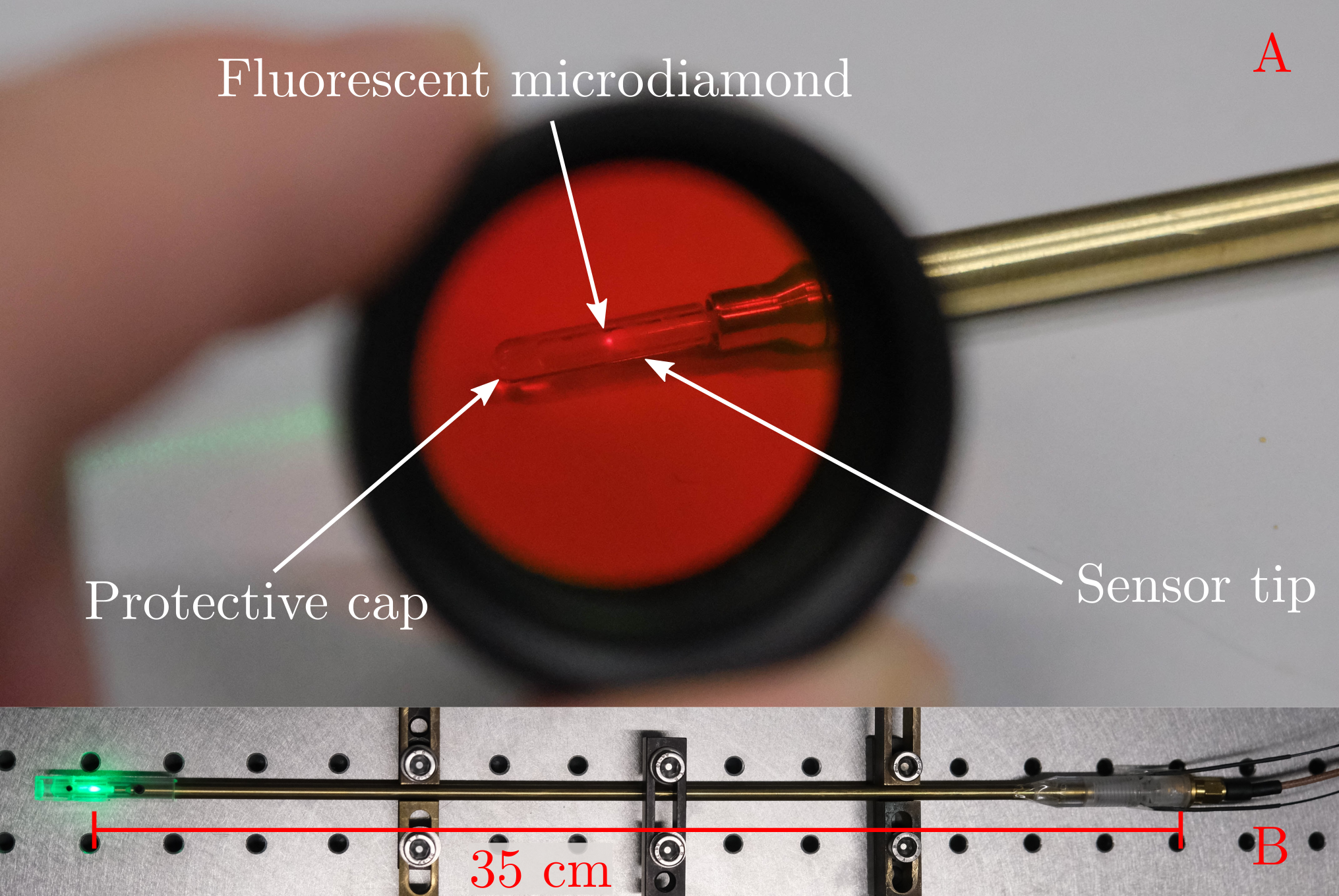}
    \caption{Images of the assembled sensor with a total length of $\approx \SI{35}{\centi\meter}$. (A) Close view of the sensor tip encased in a brass rod with an additional protective cap. The fluorescence of the diamond during excitation through the multi-mode fiber is shown through a longpass filter ($>\SI{600}{\nano\meter}$). (B) Entire sensor, including the additional optional cap for a constant magnetic bias field, the optical fibers, and the SMA connector for the MW signal connection at the back of the sensor.}
    \label{fig:sensor_komplett}
\end{figure}

The fabrication of the sensor tip starts by creating a silver antenna structure using a photo-activated reduction process that has already been introduced in \cite{dix_fiber-tip_2022}.
\begin{figure}[htbp]
    \centering
    \includegraphics[width=0.67\linewidth]{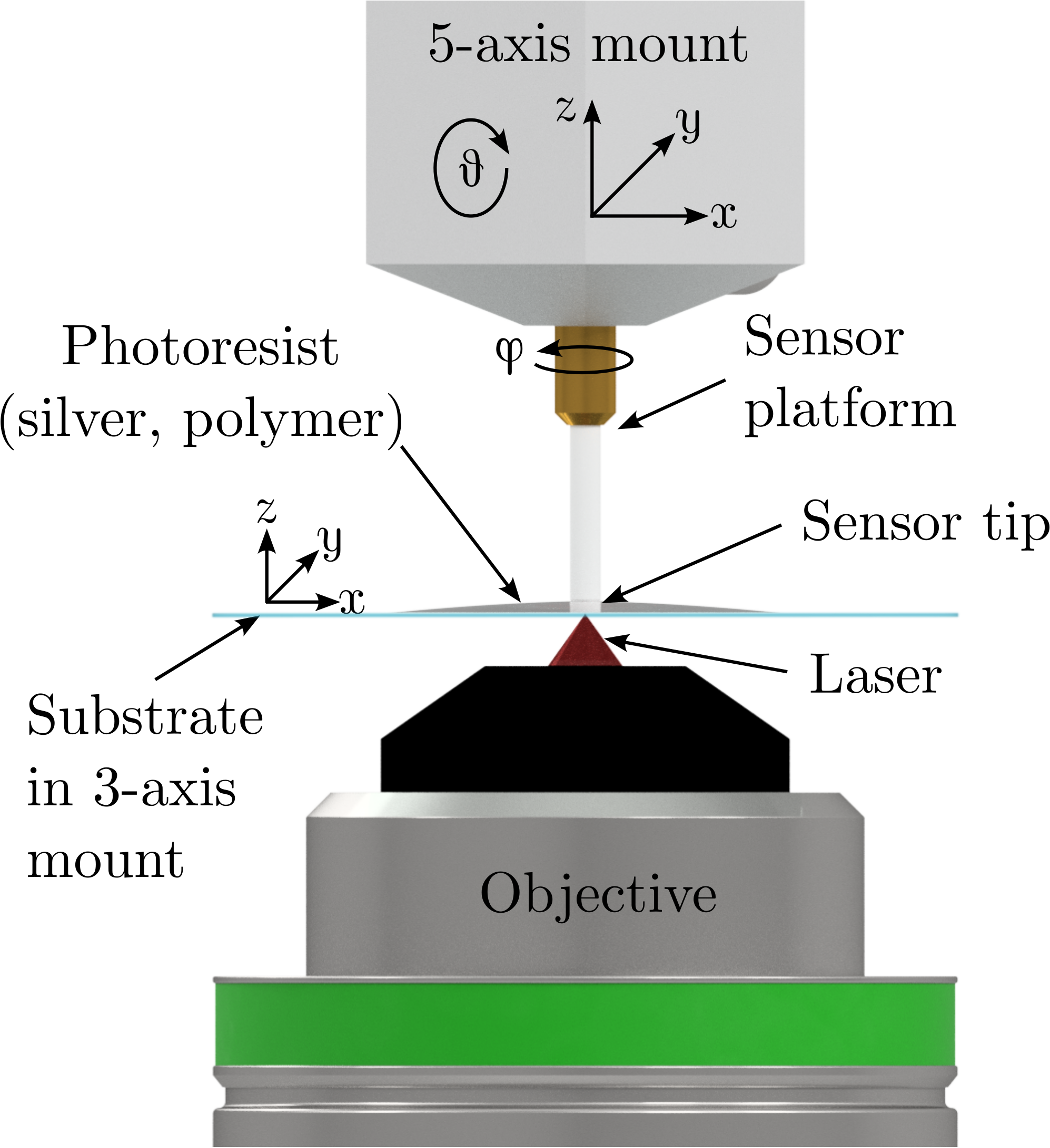}
    \caption{
        Mounted sensor platform during the direct laser writing process. The resist used is either a water-based silver solution or a polymer resist with suspended diamonds.
    }
    \label{fig:dlw}
\end{figure}
Therefore, the sensor platform is positioned as shown in FIG. \ref{fig:dlw}.
For the silver laser writing, additionally, a small reservoir filled with a silver resist encompasses the sensors' tip.
Using a 2D silver resist \cite{dix_fiber-tip_2022, waller_functional_2019} and a $\SI{20}{}\mathrm{x}$ $0.5$ NA objective (EC Epiplan-Neofluar, Zeiss), we directly write a silver microstrip-like antenna with a width of $\SI{10}{\micro\meter}$ and thickness of $\SI{5}{\micro\meter}$ as shown in FIG. \ref{fig:assemblysensor} (B).
After rinsing the tip with water to remove unprocessed resist, the structure can already be used to excite the microwave transitions of NV centers without further processing.
Optionally, a subsequent galvanization step can be conducted after the silver DLW process, which increases the efficiency of the MW emission and lowers the ohmic resistance of the antenna by approximately one order of magnitude.
For the sensor shown in FIG. \ref{fig:sensor_komplett} we galvanized the structure for $\SI{2}{\minute}$ at $\SI{0.7}{\volt}$ using a commercial silver solution, which reduced the ohmic resistance from $\approx\SI{35\pm 0.5}{\ohm}$ to $\approx\SI{3.3\pm 0.5}{\ohm}$ when measured at the SMA (SubMiniature version A) connector, which is soldered to the end of the silver wires.

In order to connect a diamond containing NV centers to an optical fiber, the tip of the multi-mode (MM) fiber is positioned $\SI{50}{\micro\meter}$ above the diamond using an external 5-axis stage, and the stages and imaging system of the microscope of the Nanoscribe system as depicted in FIG. \ref{fig:dlw}.
We use a polymer resist (IP Visio), which provides a low fluorescence and high optical transparency in the visible spectrum \cite{schmid_optical_2019}.
As a connecting and light-guiding element, a cone structure is direct laser written between the core of the MM fiber and the diamond.
The base of the cone is adapted to the size of the core of the MM fiber, and the tip of the cone is designed to be slightly larger than the diamond.
Due to the scattering of the laser beam in the center of the cone caused by the microdiamond, the structure is written hollow using only contour lines.
A second polymer structure is written between a single-mode (SM) fiber and the diamond to connect the SM fiber to the diamond for an additional optical excitation beam path.
It is shaped such that optical excitation light hits the diamond from the side, reducing the overall excitation light in the MM fiber, as depicted in FIG. \ref{fig:assemblysensor} (C, D).
The written structure is post-cured using a UV LED after the chemical development.
Further details can be found in Appendix \ref{sec:appendix A}. 

As a last step, the sensor is mounted in an optional metallic or plastic tube with a $\SI{5}{\milli\meter}$ outer and $\SI{3}{\milli\meter}$ inner diameter to enhance mechanical stability, as shown in FIG. \ref{fig:sensor_komplett} (A). 
The antenna's resonance is tuned by measuring the $\mathrm{S_{11}}$ parameter and iterative shortening of the silver wires until a resonance of the $\mathrm{S_{11}}$ parameter is measured at $\SI{2.8\pm0.2}{\giga\hertz}$. 
An image of the assembled sensor is shown in FIG. \ref{fig:sensor_komplett} (B). 
The distance between the MW connector and the tip is $\approx \SI{35}{\centi\meter}$, while the optical fibers have a length of $\approx \SI{1.7}{\meter}$, making it an easy-to-mount-and-hold but flexible sensor.

\begin{figure*}[htbp]
    \centering
    \includegraphics[width=\textwidth]{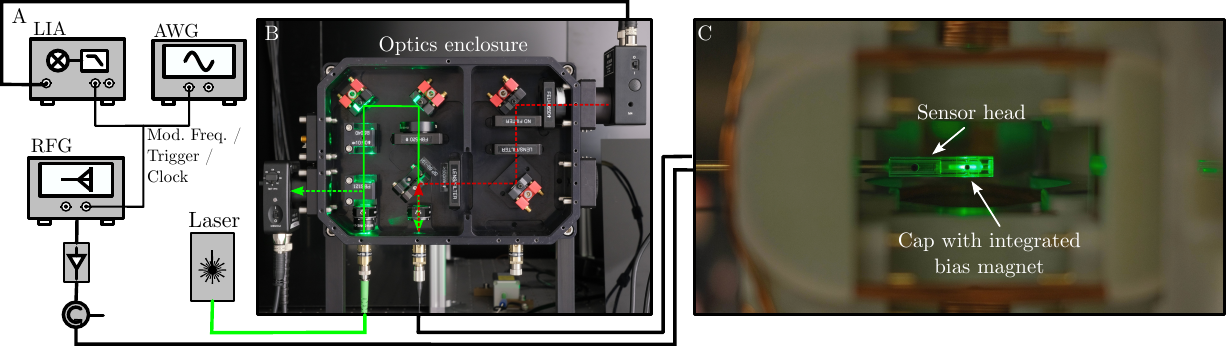}
    \caption{
        Schematics of the electrical and optical setup, specifically in a configuration for LIA measurements.
        (A) The MW signal is frequency modulated and directly connected to the sensor. The fluorescence signal, measured either with a photodiode (PD) or an avalanche PD (APD), is demodulated and filtered by a LIA.
        (B) Optical fiber coupling is done in a custom anodised enclosure, milled from aluminium.
        A first half-wave plate and a polarizing beam cube ensure a linear polarization of the in-coupled laser light and allow monitoring changes in laser power caused by fluctuations in the polarization via a PD.
        A second non-polarizing beam cube allows monitoring of the optical excitation power and can be used for further improvements using balanced detection schemes.
        ND and laser clean-up filters can be inserted, if needed, before the laser light is coupled into the sensor head fiber. 
        The collected fluorescence is either separated by a dichroic mirror for single fiber measurements or, for dual fiber measurements, fluorescence collected from the sensor head is guided back through the second fiber into the optical setup and then filtered with a longpass filter.
        Finally, the fluorescence intensity is either measured by a PD or an APD or coupled to a MM fiber which guides the fluorescence light to a SPCM.
        (C) The sensor head inside of the coil system. A bias magnetic field can be applied with a screw-on cap containing an integrated magnet.
    }
    \label{fig:setup}
\end{figure*}

\section{Results}\label{sec:results}

The sensor's design is characterized by its potential to reduce autofluorescence in the detecting optical fiber and its sensitivity. The optics enclosure, depicted in FIG. \ref{fig:setup}, is utilized for all measurements. A detailed description of the optical setup is given in Appendix \ref{sec:setup}.

\subsection{Fluorescence measurements}\label{sec:autofluorescence}
Before characterizing the sensitivity of our sensor, we measure its relevant fluorescence properties and the reduction of the autofluorescence due to the use of the dual-fiber sensor, as depicted in FIG. \ref{fig:assemblysensor} (D).
For this purpose, first, the transmitted laser power in the detecting MM fiber in the dual-fiber configuration with the attached structure and diamond is measured and compared to the laser power  coupled into the SM fiber.
Next, the autofluorescence intensity at a given laser power in an identical unprocessed MM fiber without an attached diamond and the sensor's fluorescence in the single-fiber configuration is measured.
We then calculate the reduction in laser power by comparing the amount of laser power guided in the detecting MM fiber in both configurations.
Assuming a linear scaling of the autofluorescence with the laser power, the potential reduction in autofluorescence, which arises from the transmitted light, is estimated.
From this ratio, we calculate the minimal possible diamond size in a dual- and single-fiber-based sensor, where the fluorescence of the diamond is not drowned out by the autofluorescence of the fiber.

For the first measurement, we detect a transmitted laser power through the detecting MM fiber in the dual-fiber configuration of $\SI{2.04}{\micro\watt}$ at a laser power transmitted through the SM fiber of $\SI{1.64}{\milli\watt}$.
Thus, we find a reduction of the laser light in the detection fiber by a factor of $\SI{804}{}$.
Next, we measure the amount of autofluorescence and fluorescence transmitted through a longpass filter with a cutoff wavelength of $\SI{650}{\nano\meter}$ and neutral density (ND) filters using a single photon counting module (SPCM).
The absolute count rate is calculated by applying correction factors for each ND filter at the given transmission value at a wavelength of $\SI{685}{\nano\meter}$, close to the NV center's maximum emission wavelength of $\SI{684}{\nano\meter}$ at room temperature \cite{sanchez_toural_hands-quantum_2023}.
We detect an autofluorescence intensity of $\SI{3.9E8}{}\,\mathrm{cps}$ (counts per second) from the unprocessed MM fiber and $\SI{3.3E12}{}\,\mathrm{cps}$ of fluorescence light from the $\SI{15}{\micro\meter}$-sized diamond at an excitation laser power of $\SI{2.2}{\milli\watt}$ in the single-fiber configuration.
This yields a contrast ratio of $\approx \frac{8410}{1}$, from which we estimate the minimal diamond size necessary to have a contrast ratio $\mathrm{SNR}=1$.
Hence, we find a minimum size of $\SI{740}{\nano\meter}$ when assuming a constant laser power and a constant NV doping in a single-fiber configuration.
In comparison, integrating much smaller diamonds on the tip of a fiber down to a size of $\SI{79}{\nano\meter}$ and the same NV-doping of ($\approx\SI{3.5}{}\,\mathrm{ppm}$) is possible with a dual fiber configuration, enabling even higher spatial resolution in the same sensor package.
However, we also find a reduction in the fluorescence intensity by a factor of $\SI{9.9}{}$ in the dual-fiber configuration.
We attribute this reduction to the worse coupling efficiency of the single-mode fiber and limitations in the fabrication process, where the used air objective provides an elongated focus shape, resulting in an imperfect, elongated waveguide structure. 
In summary, the dual-fiber-based approach overcomes elementary issues associated with autofluorescence when using smaller diamond sizes, which can be realized using the same setup. 
However, the losses due to the imperfect waveguide structure significantly reduce the amount of fluorescence.
Consequently, we achieve better sensitivities in the single-fiber configuration, knowing that the SNR is only moderately affected when utilizing a $\SI{15}{\micro\meter}$-sized diamond and the fluorescence, due to the more efficient usage of laser power, is $\approx 10$ times higher.
Therefore, if not stated otherwise, the following measurements are conducted in a single-fiber configuration.
For smaller diamonds, however, physical separation of the light path for excitation and detection yields significant advantages, and technical limitations can be overcome by using immersion objectives with higher numerical apertures. 
Results of a similar structure with a similar resist (OrmoComp, micro resist technology), a slightly modified design that compensates for the physical elongation, and a $25\times$ $0.8$ NA immersion objective yield an identical efficiency in the optical excitation in both configurations. Therefore, we expect that the result is mainly limited by the imperfect structure.

\subsection{Sensitivity characterization}\label{sec:sensitivity}
\begin{figure}[htbp]
    \centering
    \includegraphics[width=\linewidth]{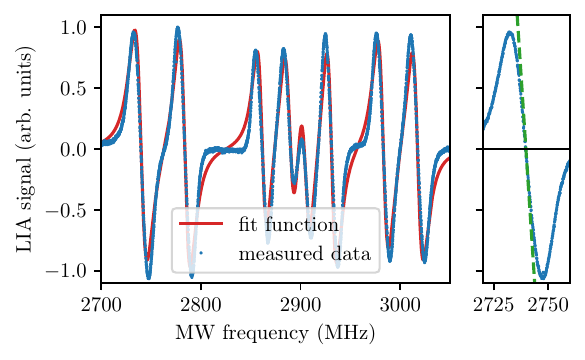}
    \caption{
        Left: Lock-in amplified ODMR signal while a bias magnetic field is applied (solid blue).
        Resonances are fitted with derivatives of Lorentzian functions (red), where the zero-crossing marks the resonance frequency.
        Right: The magnetic-field sensitivity is dependent on the calculated slope at the zero-crossing point that is visualized as a dashed green line.
        }
    \label{fig:odmr_lia}
\end{figure}

Sweeping over a range of MW frequencies results in an ODMR spectrum of the NV resonances, as shown in FIG. \ref{fig:odmr_lia}.
We track the $X$ component of the LIA output over time, and the MW frequencies are calculated via the sweeping time.

An unconstrained non-linear least squares fit method (Levenberg-Marquardt) \cite{watson_levenberg-marquardt_1978} is used to fit a model function of a sum of the derivatives of eight Lorentzian distributions
\begin{equation}
    L(f) = \sum_{i=1}^8 -\frac{32\sqrt{3}}{9} \frac{C_iw_i^3(f-f_i)}{(4(f-f_i)^2+w_i^2)^2},
\end{equation}
where for the $i$-th resonance ($i\in(1,...,8)$) $C_i$ is the peak contrast, defined as the maximum value, $f_i$ is the resonance frequency, and $w_i$ is the FWHM linewidth.
We found that to automate the selection of starting points for the fit model, using the mean abscissas between the peaks and valleys for each resonance frequency leads to reasonably stable fitting behavior.

The achieved magnetic-field sensitivity is highest at the zero-crossing point, where the fit function has a maximal slope $S_{i} = \frac{32\sqrt{3}}{9} \frac{C_i}{w_i}$.
A small change in the external magnetic field along the NV-axis will shift the resonances by the gyromagnetic ratio $\gamma = g_e\mu_B/h \approx \SI{28.03}{\giga\hertz\per\tesla}$ of the NV center \cite{rondin_magnetometry_2014}.
The output of the LIA will then react with a change of output voltage at the previous resonance frequency of magnitude
\begin{equation}
    R_i = \gamma S_i = \frac{32\sqrt{3}}{9}\frac{g_e \mu_B}{h}\frac{C_i}{w_i},
\end{equation}
where $R$ is called the magnetometer response.
The magnetic-field sensitivity, defined as the minimal change of magnetic field $\delta B$ that can be resolved in a given time $t$, is related to the standard deviation of the LIA output $\sigma$, the magnetometer response $R$ and the noise-equivalent power (NEP) bandwidth of the LIA's low-pass filter $f_\text{NEP}$ \cite{schloss_simultaneous_2018}
\begin{equation}
    \eta_{\text{LIA}} = \delta B \sqrt{t} = \frac{\sigma}{R\sqrt{2f_{\text{NEP}}}} \left[\SI{}{\tesla\per\sqrt{\hertz}}\right].
\end{equation}
We reach an optimal sensitivity of $\SI{51.8}{\nano\tesla\per\sqrt{\hertz}}$ for one continuous ODMR sweep for a MW power of $\SI{50}{\milli\watt}$ and laser power of $\SI{2.15}{\milli\watt}$.
A more detailed discussion about optimizing measurement parameters can be found in Appendix \ref{sec:optimizations}.

\begin{figure}[htbp]
    \centering
    \includegraphics[width=\linewidth]{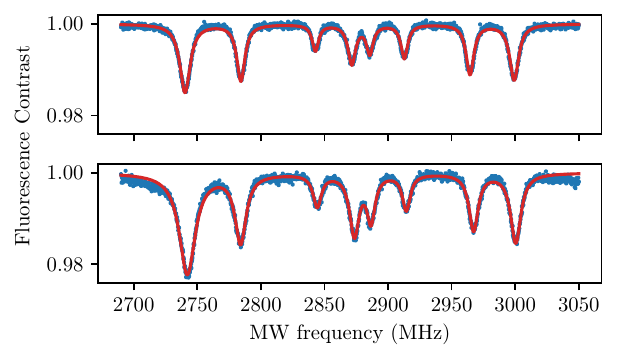}
    \caption{
        The upper graph shows the ODMR spectrum, when exciting and detecting via the MM fiber of the sensor. The lower graph shows the ODMR spectrum, when exciting via the SM fiber and detecting through the MM fiber.
        Resonances are fitted with Lorentzian functions (red) to obtain contrast and linewidth.
    }
    \label{fig:odmr_spcm}
\end{figure}

To estimate the photon shot-noise limit, we average over multiple ODMR measurements with a SPCM at the same MW and laser powers and calculate the shot-noise-limited sensitivity for each axis according to \cite{dreau_avoiding_2011}
\begin{equation}
    \eta_{\text{SN}} = \frac{4}{3\sqrt{3}} \frac{h}{g_e\mu_B} \frac{w}{C\sqrt{I_0}} \left[\SI{}{\tesla\per\sqrt{\hertz}}\right],
\end{equation}
where $I_0$ is the photon collection rate, corrected for optical losses due to the filters and the detectors efficiency.
The most optimal shot-noise-limited sensitivity for one resonance is $\SI{5.9}{\nano\tesla\per\sqrt{\hertz}}$.
To test our two-fiber scheme for reducing the impact of fiber autofluorescence, we compare the sensitivity to an otherwise identical measurement but excite the NV centers via the SM fiber and still detect the signal using the MM fiber.
The ODMR spectra, in which we optically excite the NV centers via the MM and SM fiber, are shown in FIG. \ref{fig:odmr_spcm}. 
The sensitivity of all NV axes is listed in Tab. \ref{tab:sensitivities}.
The best shot-noise-limited sensitivity of our sensor is achieved when exciting and detecting through the MM fiber ($\SI{5.9}{\nano\tesla\per\sqrt{\hertz}}$) compared to the identical measurement with excitation through the SM fiber ($\SI{27.7}{\nano\tesla\per\sqrt{\hertz}}$).
We attribute this decrease in sensitivity to a reduced fluorescence due to a lack of optical excitation power from the SM fiber and MW-induced power broadening of the linewidth.
Normalizing the sensitivity with respect to the volume of the diamond enables comparability with other sensors and, thus, the characterization of the efficiency of the sensor platform independent of the size of the diamond. 
Therefore, we multiply the sensitivity by the square root of the diamond size or the sensor volume, as the sensitivity increases with the square root of the number of emitting spins.
The volumetric SN-limited sensitivity related to our total sensor head size is $\SI{42.9}{{\milli\meter}^{3/2} \times  \nano\tesla \per \sqrt{\hertz}}$ and to the diamond size is $\SI{7.8}{ {\milli\meter}^{3/2} \times \pico\tesla \per \sqrt{\hertz}}$ assuming a homogeneous $\SI{15}{\micro\meter}$-sized sphere for the diamond.

\subsection{Noise floor}\label{sec:noise}
Next, we characterize the noise floor of our sensor by keeping the MW excitation frequency fixed to one resonance and evaluating the amplitude spectral noise density (SND) \cite{sekiguchi_diamond_2024, webb_nanotesla_2019, ahmadi_pump-enhanced_2017} and the overlapping Allan deviation \cite{allan_statistics_1966}.

\begin{figure}[htbp]
    \centering
    \includegraphics[width=\linewidth]{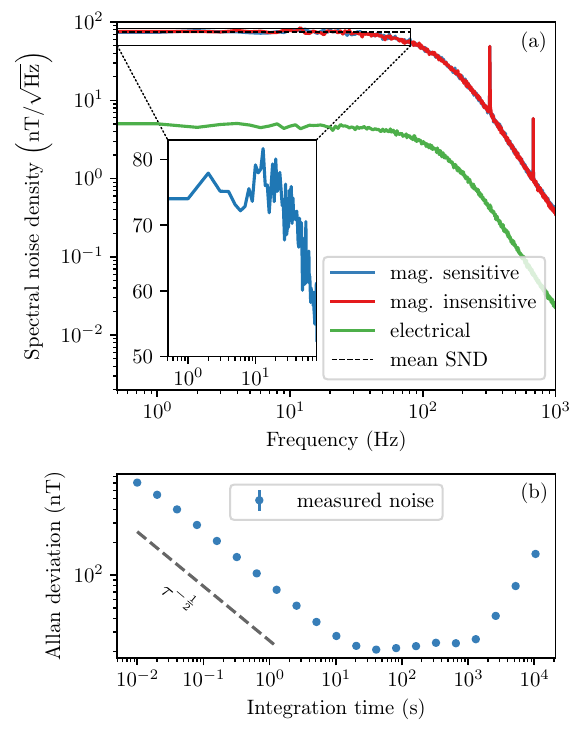}
    \caption{
        (a) Amplitude spectral noise density of our sensor for magnetically sensitive (blue), magnetically insensitive (red), and electrical (green) noise measurements.
        We reach a mean spectral noise density of $\SI{74.6}{\nano\tesla\per\sqrt{\hertz}}$ (dashed black) in a $\SI{3}{\deci\bel}$ bandwidth of $\SI{80.9}{\hertz}$.
        (b) Overlapping Allan deviation of our sensor.
        Errorbars for the Allan deviation are to small to be visible.
    }
    \label{fig:fft}
\end{figure}

For the SND, the resulting signal is binned into windows of $\SI{1}{\s}$ duration each.
We then calculate the absolute value of the averaged single-sided discrete Fourier transform $|\mathcal{F}|$ of the signal.
The SND follows from the relation \cite{patel_subnanotesla_2020, van_baak_response_2014}
\begin{equation}
    \text{SND} = \frac{|\mathcal{F}|}{R} \sqrt{\tau} \left[\SI{}{\tesla\per\sqrt{\hertz}}\right],
\end{equation}
where the Fourier transform is scaled by the magnetometer response $R$ and the single-sided bandwidth $1/\tau$ of the rectangle window function, which results from the binning of the discrete Fourier transform.
The bandwidth scaling is proportional to $\sqrt{\tau}$, because we consider the amplitude SND, which is the square-root of the power SND.
We note that in some literature amplitude is defined as root-mean squared amplitude \cite{zhang_pulsed_2022, van_baak_response_2014}, leading to a scaling difference of $1/\sqrt{2}$ in SND.
When recording the $X$ component of the LIA output signal, the signal is said to be magnetically sensitive, while recording the $Y$ component, which is not dependent on magnetic field changes, denotes the magnetically insensitive case.
Additionally, we measure the noise without laser excitation, resulting in only electrical noise from the PD and the LIA \cite{patel_subnanotesla_2020}.
FIG. \ref{fig:fft} shows the SND for all three cases.
We notice uncancelled noise peaks at multiples of $\approx \SI{300}{\hertz}$. Because these peaks appear in both magnetically sensitive and insensitive cases, we attribute this noise to our laser. In the future, we plan to employ balanced-photo detection schemes \cite{schloss_simultaneous_2018} to further mitigate laser noise.

The bandwidth of our sensor is limited by the $\SI{3}{\deci\bel}$ bandwidth of the low-pass filter of our LIA to $\SI{80.9}{\hertz}$, as indicated by the inset in FIG. \ref{fig:fft} (a).
We reach a mean SND of $\SI{74.6}{\nano\tesla\per\sqrt{\hertz}}$ in this bandwidth.
When AC magnetic fields with higher bandwidths are to be recorded, the bandwidth of our sensor can be extended up to the modulation frequency of the LIA.
The modulation frequency can then also be further extended to higher frequencies \cite{schloss_simultaneous_2018}. 
However, a trade-off between modulation frequency and sensitivity is then necessary, as discussed in Appendix \ref{sec:optimizations}. 

To characterize the long-term stability of our sensor, we record a data set for a longer time of $\approx \SI{15}{\hour}$ and calculate its overlapping Allan deviation.
We reach a minimal sensitivity of $\SI{21.3}{\nano\tesla}$ for an integration time of $\SI{41}{\second}$ as shown in FIG. \ref{fig:fft} (b).
We conclude, that the integration of optical fibers, MW antenna, and the NV doped diamond in a fixed package on our sensor head results in a highly stable sensor platform, that is capable of resolving magnetic fields down to the $\SI{}{\nano\tesla}$ range.

\subsection{Magnetic field measurements}\label{sec:bfield}
We demonstrate a magnetic field measurement of a complex 3D Anti-Helmholtz coil system to illustrate the real-world benefits that NV-based sensors bring over conventional solutions like Hall sensors.

\begin{figure}[htbp]
    \centering
    \includegraphics[width=\linewidth]{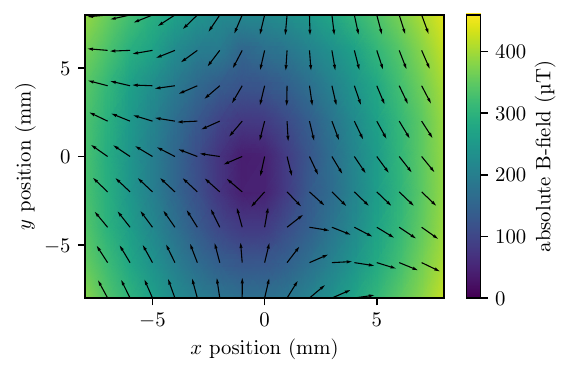}
    \caption{
        Contour plot of the interpolated measured Helmholtz coil magnetic field in Anti-Helmholtz configuration.
        Absolute field values are shown as a colorbar.
        $x$ and $y$ components of the field vector are shown as black arrows, and the base of the arrow represents the measured positions.
        }
    \label{fig:contour}
\end{figure}

The vectorial sensing capability of our sensor allows us to measure the direction of the magnetic field directly in a single measurement without the need to measure in multiple directions compared to a Hall sensor with a single Hall element. 
In our implementation, the resolution is given by the sensing volume of the NV diamond $\SI{15}{\micro\meter}$.
Here, we use spatial resolution in the intrinsic sense, i.e., the smallest spatial variation of the magnetic field that can be resolved at the sensor location. 
While the effective resolution in planar magnetic structures also depends on sensor-sample distance, for volumetric fields the intrinsic limit set by the diamond size is dominant.
For this, our sensor is fastened to a stage and moved in three dimensions through the magnetic field, and a screw-on cap with integrated permanent magnets is fixed to the tip of our sensor to apply a bias magnetic field of $\SI{10.3}{\milli\tesla}$ as depicted in FIG. \ref{fig:setup}.
Triangle-shaped sweeps of the microwave frequency at each position of the translation stages were performed to mitigate the influence of delays between the sweep and the data recording, and the following resonance frequency calculations are always averaged between one up-scan and one down-scan.
We then calculate the absolute value of the magnetic-field vector $B$ for each NV-axis $j$ from its upper and lower resonance frequencies $f_{u,j}$ and $f_{l,j}$ according to Appendix \ref{sec:b_field_derivation}.
To calibrate the sensor with the applied bias field, we calculate one magnetic-field vector without a coil magnetic field. 
This bias field is then subtracted from subsequent measured field vectors.
FIG. \ref{fig:contour} shows one scan of magnetic-field vectors in one $xy$-plane of many slices in z-direction inside the 3D coil system.
In conclusion, the experiment demonstrates the necessity of a robust and movable sensor in three dimensions with vectorial field sensing capability in a volumetric field, which benefits from the small size of the sensing NV-diamond and thus achieves high spatial resolution.

\subsection{Comparison of sensor designs}\label{subsec:comparison}

\begin{figure*}[htbp]
    \centering
    \includegraphics[width=\linewidth]{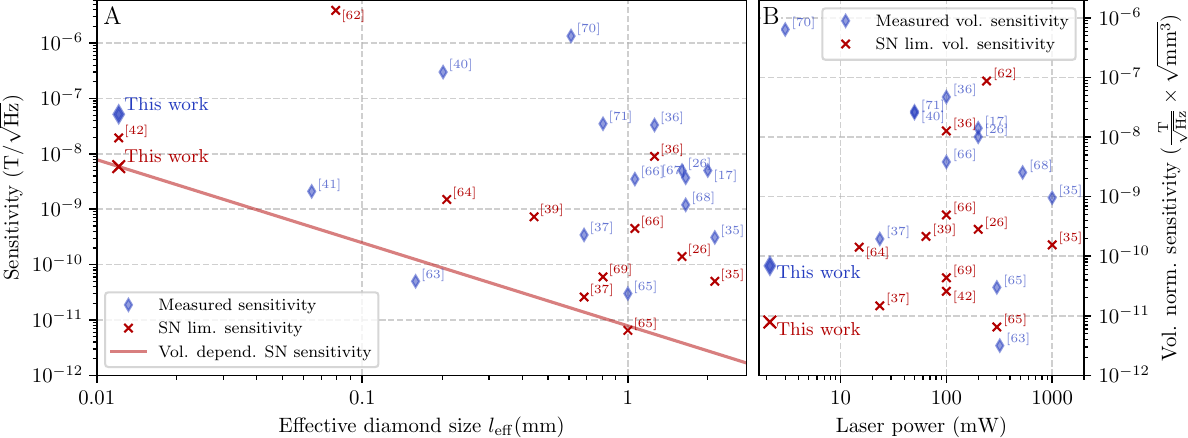}
    \caption{
        Measured and shot-noise-limited calculated sensitivities of our and other publications \cite{quan_fiber-coupled_2023, zhang_high-sensitivity_2022, dmitriev_concept_2016, graham_fiber-coupled_2023, hatano_simultaneous_2021, liu_closed-loop_2024, liu_millimeterscale_2023, blakley_room-temperature_2015, deguchi_compact_2023, blakley_fiber-optic_2016} for different effective diamond sizes (A). The values for measured sensitivities represent retrieved values from noise and time-trace measurements compared to shot-noise-limited sensitivities, which are calculated from measured contrast and linewidth of fit data and photon count rate.
        Normalized sensitivity of the presented sensors by the individual diamond volume at the given laser power (B).
        The bandwidth of the presented sensors is set differently in most cited publications and therefore not compared.
    }
    \label{fig:comparison_sens}
\end{figure*}

To provide a better overview, we compare other fiber-based sensors in the field to our approach in FIG. \ref{fig:comparison_sens} (A) in terms of sensitivity and effective diamond size $l_{\mathrm{eff}}$, which we calculate as the cube root of the given diamond volume. 
In the following, we use $l_{\mathrm{eff}}$ as a measure for the intrinsic spatial resolution of the sensing element.
Alternatively, one could consider the excited volume of the diamond, often given as the effective sensing volume, as the full diamond volume is not always excited.
We also note that the effective resolution in a given experiment additionally depends on the type of sample, e.g., when planar quasi 2D magnetic surfaces need to be mapped, the sample-sensor distance needs to be taken into account.
However, for volumetric magnetic fields inside of three-dimensional coil systems, the effective resolution is limited by the size of the sensing element.
This effective volume is not always given, and technological advantages also lie in efficiently attaching a diamond to a fiber.
Therefore, we compare the sensitivity to the diamond size in FIG. \ref{fig:comparison_sens} and, for completeness, contribute the comparison for the given effective sensing volume in FIG. \ref{fig:comparison_sens_vol} in Appendix \ref{sec:appendix E}. 
For our sensor, we consider the diamond as fully illuminated.
In FIG. \ref{fig:comparison_sens} and FIG. \ref{fig:comparison_sens_vol}, we plot the values for shot-noise-limited and measured sensitivities that are obtained using lock-in techniques and noise and time-trace measurements.
Both values, SN-limited and measured sensitivities, are important metrics.
The SN-limited sensitivity has the advantage of being independent of noise sources, e.g., electronic or magnetic field background noise. 
Therefore, the SN-limited sensitivity specifies the sensor's capabilities independent of the environment.
In contrast, the measured sensitivity is the value that represents the capabilities in a given technical setup and environment, including the existing background.
Furthermore, many publications provide only shot-noise-limited sensitivity values, while others present only the measured sensitivity.
Assuming the sensitivity evolves proportional to $\nicefrac{1}{\sqrt{N}}$, where N is the number of spins and N scales proportional to the diamond volume, we include a line that shows the expected sensitivity if one could increase the diamond size, ignoring other effects such as coupling efficiency, field and diamond inhomogeneities, collective NV-effects, and scattering. 
In FIG. \ref{fig:comparison_sens} (A), a sensor with a sensitivity below the red line could be achieved by a more efficient fluorescent light collection, a brighter diamond, a higher fluorescence contrast, or a smaller linewidth. 
As shown, only a few publications can reach this line, yet the presented ones utilize externally attached MW antennas or $^{12}\mathrm{C}$-enriched diamonds, tracking pulse techniques, and use larger diamonds. 
Furthermore, in FIG. \ref{fig:comparison_sens} (B), we plot the normalized sensitivities of the presented sensors by the respective diamond volume for the given laser powers. 
Here, our sensor provides one of the lowest volume normalized sensitivities at the lowest laser power, making it especially suitable for light-sensitive applications.

\section{Conclusion and Outlook}\label{sec:conclusion}

The miniaturized fiber-based sensor we present here provides a small sensor platform capable of measuring magnetic fields in full vectorial resolution.
The direct laser writing of polymer and silver structures onto the sensor's tip enables us to fully integrate the diamond and a microwave antenna onto the sensor head.
Together with a diamond size, and thus spatial resolution of $\SI{15}{\micro\meter}$, we have demonstrated a highly sensitive, small, and robust sensor platform and show an integrated fiberized NV-based sensor with a single diamond with the smallest diamond size published to date.
In single-fiber configuration, we achieve a shot-noise limited sensitivity of $\SI{5.9}{\nano\tesla\per\sqrt{\hertz}}$ while only using $\SI{2.15}{\milli\watt}$ of laser power and $\SI{50}{\milli\watt}$ of microwave power.
The sensitivity of the sensor using lock-in-amplification techniques is $\SI{51.8}{\nano\tesla\per\sqrt{\hertz}}$ at the same powers and configuration.

When analyzing the fluorescence properties of our sensor, we find that with our approach, we can reduce the amount of autofluorescence in the detecting fiber by a factor of $\approx \SI{800}{}$ and, thus enabling the integration of even smaller diamonds with only a small impact on the SNR.
We emphasize that losses in laser light and slightly worse sensitivity in the dual-fiber configuration result from an imperfect waveguide structure. 
The used air objective provides an elongated voxel shape and thus elongated structure, significantly affecting light-guiding properties. 
Sensors with an optimized structure using an immersion medium-matched objective with a higher NA exhibit an identical excitation efficiency independent of the fiber used for excitation.
Using our sensor, we show, as a practical use case, the measurement of vectorial magnetic fields of a coil system.
Furthermore, we emphasize that our approach shows that reducing the total sensor size can achieve high sensitivities at low laser and microwave powers, enabling light and microwave-sensitive measurements.

The presented platform provides a physically robust solution for the sensor head with flexible cables and fibers that allows the integration of diamonds of varying sizes and a high sensitivity. The dual-fiber configuration, while critical to manufacture precisely, allows the integration of diamonds of varying sizes and can overcome the autofluorescence limitations of single-fiber sensors down to a diamond size below $\SI{100}{\nano\meter}$.
Integration of diamonds below the optical resolution of the microscope system is possible. 
However, further techniques are needed to identify single diamonds, e.g., by ODMR measurements of a sample with a magnetic field applied just before attaching a diamond to the sensor tip.

In addition to CW ODMR measurements shown here, our sensor platform is compatible with the full range of pulsed magnetometry schemes. 
The technical capabilities to pulse the laser and MW excitation can be added independently of the sensor platform by, for example, switching the laser diode current and the MW signal with a PIN diode. 
Pulsing laser and MW excitation would allow the implementation of all known coherent spin control sequences like Rabi, Ramsey, or Spin-Echo measurements. 
Such pulsed measurements then allow for the measurement of not only DC magnetic fields but also AC magnetic-field measurements. 
Pulsed measurement sequences can further improve the magnetic-field sensitivity by suppressing specific noise sources. 

In conclusion, we present a reliable, reproducible sensor platform for fiberized fluorescent particle sensors at small scales with a high volume normalized sensitivity. 
We use functional photonic waveguide polymer and silver-antenna structures on the tip of optical fibers and electrical wires and show an approach to overcome existing limitations of autofluorescence. 
The presented sensor platform and techniques also pave the way for future developments of advanced fiber-based sensors that could potentially integrate multiple diamonds with varying sizes and networks combined with advanced photonic structures.

\section{Data and code availability}
All data shown in the images, plots and data supporting the findings in this publication is available on Zenodo \cite{johansson_data_2024}. The code library used to control the devices for all measurements can be found in \cite{noauthor_microscope_2024}.

\section{Acknowledgements}
We acknowledge support by the nano-structuring center NSC.
This project was funded by the Deutsche Forschungsgemeinschaft (DFG, German Research Foundation) Project-ID No. 454931666, the German Federal Ministry of Education and Research (BMBF) as part of the QuanTEAM project (FKZ13N16467), and the Quanten-Initiative Rheinland-Pfalz (QUIP).
Furthermore, we thank Dr. Erik Hagen Waller for providing the silver resist as well as Prof. Dr. Georg v. Freymann, Prof. Dr. Elke Neu-Ruffing, Alexander Bukschat, Nimba Pandey, Oliver Opaluch and Tobias Pätkau from the RPTU University Kaiserslautern-Landau for helpful discussions and experimental support. 
We also thank Dr. Gesa Welker and Dr. Samer Kurdi from the van der Sar Lab at TU Delft for their support on technical questions.

\appendix

\section{Geometrical design of the polymer structure}\label{sec:appendix A}
\begin{figure}[htbp]
    \centering
    \includegraphics[width=\linewidth]{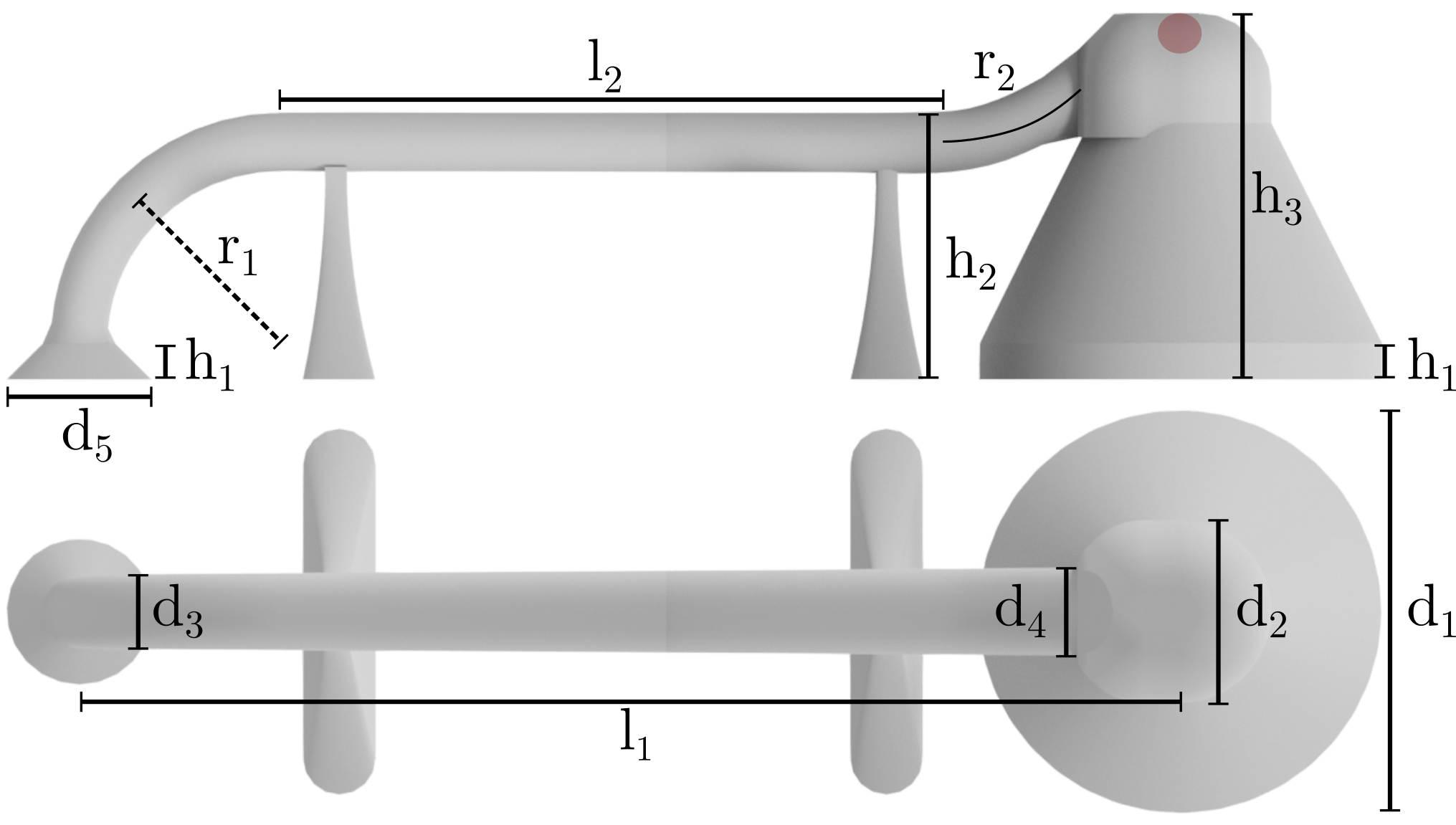}
    \caption{Schematic rendering of the designed polymer structure and given geometries. The red dot indicates the intended position of the microdiamond.}
    \label{fig:Polymerstructure}
\end{figure}

A schematic image of the designed polymer waveguide structure is shown in FIG. \ref{fig:Polymerstructure}.

The geometrical shape of the polymer structure is motivated by a set of functional and fabrication-related constraints. 
The primary objective is to fix a diamond above the core of a multi-mode fiber while ensuring efficient light guidance between the fiber and the diamond. Additionally, the structure must facilitate efficient coupling between a single-mode fiber and the diamond while minimizing coupling into the multi-mode fiber core. The width and shape of the polymer structure at the connection between the fiber core and the polymer structures need to be matched in diameter such that most light is guided from the fiber to the diamond. 
Furthermore, the structure has to be optimized for strength without interfering with the antenna or waveguide. 
To ensure structural integrity during the undocking process after direct laser writing, adhesion should be strong at the fiber tip and weak at the glass substrate.
Therefore, the contact area between the fiber tip and the polymer structure must be maximized without affecting the light-guiding properties, which directly impacts $d_1$, $d_5$, and supporting structure geometry.
In contrast, the contact area between the glass substrate and the polymer structure must be minimized, which is achieved by different heights ($h_2=\SI{37}{\micro\meter}$) of the light guiding waveguide structure and the attached diamond at the very tip $h_3$, as well as by the diameter of the surrounding polymer material around the diamond $d_2$.
The base of the cone has a diameter of $d_1=\SI{55}{\micro\meter}$ and is slightly larger than the $\SI{50}{\micro\meter}$ sized core of the MM fiber having ($\SI{5}{\micro\meter}$) of positional tolerance and slightly increasing the adhesion between polymer structure and optical fiber. 
Furthermore, the cone stands on top of a $h_1=\SI{5}{\micro\meter}$ tall cylinder to compensate any tilts and focus mismatching and improve adhesion. 
The tip of the cone is designed to be slightly larger ($d_2=\SI{30}{\micro\meter}$) than the diameter of the $\SI{15}{\micro\meter}$ sized diamond. 
The edges of the tip of the cone are also rounded to reduce physical contact with the glass substrate below. The total height $h_3$ is $\SI{50}{\micro\meter}$. 
Due to scattering of the laser beam in the center of the cone caused by the microdiamond during the direct laser writing process, the structure is written in 12 contour-lines with $\SI{0.4}{\micro\meter}$ spacing in between and a top layer thickness below and around the diamond of only $\SI{12}{\micro\meter}$. 
Furthermore, a structure connecting a single-mode (SM) fiber to the diamond for an additional optical excitation light-guiding beam path is added.
This waveguide structure consists of a small cone with a base diameter of ($d_5=\SI{20}{\micro\meter}$) and height of $\SI{5}{\micro\meter}$ at which the cone diameter is $d_3=\SI{10}{\micro\meter}$. 
This cone has no effect on the optical properties but increases contact area and the diameters $d_3$ and $d_4$ were set to be wide enough to fully illuminate the diamond as well as to have minimal impact on light collection efficiency of the cone above the MM fiber.
The radii of the curved waveguide structure are designed to be as large as possible, considering other boundary conditions.
The initial $\SI{10}{\micro\meter}$ wide waveguide follows a $\SI{90}{\degree}$ curve with a $r_1=\SI{28}{\micro\meter}$ radius a straight line with a length of $l_2=\SI{88}{\micro\meter}$ and is then guided lightly upwards at a $r_2=\SI{30}{\micro\meter}$ radius which then guides light centrally at the designed diamond center position.
At this position, the waveguide has a width of $d_4=\SI{12}{\micro\meter}$, which is increased gradually. 
For the chosen radii $r_1$, $r_2$, no significant change is expected based on previous work \cite{landowski_17}.
The waveguide structure is supported via two small towers and intentionally designed such that it does not bond to the glass substrate, which is at the same height as the diamond. 
Furthermore, the geometrical value of $l_1=\SI{150}{\micro\meter}$ is given by design from the diameters of fibers $d_\mathrm{fiber}=\SI{125}{\micro\meter}$, wires $d_\mathrm{wire}=\SI{100}{\micro\meter}$ and inner ferrule diameter $d_\mathrm{ferrule}=\SI{270}{\micro\meter}$.
Further details of all direct laser-written structures can be retrieved from the CAD models in the supplemental material.

\section{Measurement Setup}\label{sec:setup}

Our optical setup consists of a modular compact enclosure with a size of $\SI{215}{} \times \SI{150}{} \times \SI{58}{\cubic\milli\meter}$, which contains all necessary optical elements as well as some optional elements to connect an excitation laser, fiber-sensor, and various detectors.
A fiber-coupled laser (Cobolt 06-MLD, Hübner Photonics) with a wavelength of $\SI{520}{\nano\meter}$ is used to excite the NV centers optically.
To avoid damaging the sensor head through excessive heating, we limit the laser power to $\SI{2.2}{\milli\watt}$ in the following measurements.
The laser light is directly coupled into the enclosure, where first, a half-wave plate (WPH05ME-514, Thorlabs) and a polarizing beam cube (PBS121, Thorlabs) ensure a linear polarization of the in-coupled laser light.
The reflected part of the beam can be monitored by a photodiode (PD) and enables measuring changes in power caused by fluctuations in the polarization.
A second non-polarizing beam cube (BS040, Thorlabs) allows monitoring of the optical excitation power and can be used for further improvements using balanced detection schemes.
ND filters and laser clean-up filters can be inserted if needed before the laser light is coupled into the sensor head fiber. 
The collected fluorescence is either separated by a dichroic mirror (SP 69-191, Edmund Optics) for single fiber measurements or, for dual fiber measurements, fluorescence collected from the sensor head is guided back through the second fiber into the optical setup and then filtered with a longpass filter (FELH650, Thorlabs) and further optional ND filters.
Finally, the fluorescence intensity is either measured by a photodiode (SM05PD3A, Thorlabs) or an avalanche photodiode (APD) (APD130A, Thorlabs) or coupled to a multi-mode fiber (M43L05, Thorlabs) which guides the fluorescence light to a single photon counting module (SPCM) (COUNT-T-100-FC, LaserComponents).

The MW signal used is first generated by a radio-frequency generator (SMB100A, Rohde\&Schwarz) and then further amplified by an amplifier (ZHL-16W-43-S+, Mini-Circuits) by $\SI{46}{\deci\belm}$.
We limit the MW power in the following measurements to $\approx\SI{50}{\milli\watt}$ to avoid thermally damaging the sensor head.
However, we see significant broadening of the ODMR linewidths, indicating that our sensitivity is not limited by the MW power.
A coaxial isolator (JIC2700T3500S2, JQL Technologies) is inserted to avoid back reflections before the MW signal is guided via the SMA connector to the fiber head antenna.
To determine the actual microwave power at the sensor, we measure the absolute gain between the MW signal generator and the SMA connector of the sensor with a vector network analyzer (MS2038C, Anritsu). 
We obtain an absolute amplification of the signal in combination with all loss elements (circulator, amplifier, connectors, and connecting cable) of $\SI{38}{\deci\bel}$.

To perform lock-in amplification of the measured PD/APD signal, we sine-wave modulate the MW frequency.
The required reference signal is generated by an arbitrary waveform generator (33622A, Keysight).
The PD/APD signal is AC coupled, using a $\SI{100}{\nano\farad}$ capacitance in series, to a lock-in amplifier (LIA) (MFLI $\SI{500}{\kilo\hertz}$, Zurich Instruments), which demodulates and filters the signal.
We apply a 3rd order low-pass filter with a $\SI{1}{\milli\s}$ time constant (equivalent to a $\SI{3}{\deci\bel}$ bandwidth of $\SI{80.9}{\hertz}$ or a noise-equivalent power bandwith of $\SI{93.5}{\hertz}$) with the LIA.
The phase shift between the reference signal and the measured fluorescence signal is always chosen to maximize the signal of the $X$ component.

All devices are connected to the same $\SI{10}{\mega\hertz}$ clock reference signal.
Appendix \ref{sec:optimizations} contains a more detailed discussion about optimizing individual measurement parameters.

\section{\label{sec:optimizations}Sensitivity optimizations}
\begin{figure}[htbp]
    \centering
    \includegraphics[width=\linewidth]{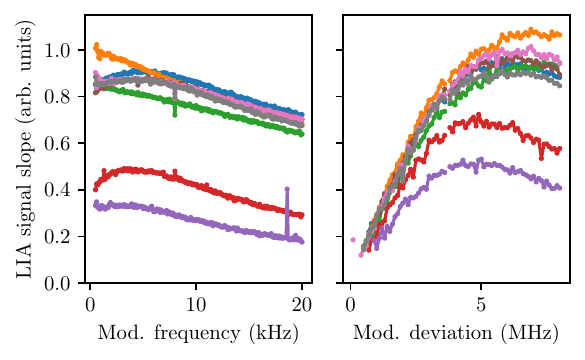}
    \caption{
        Zero-crossing slope of the LIA output for all eight resonances dependent on MW modulation frequency (left) and modulation deviation (right), each color representing one resonance.
        Each resonance generally has a different set of parameters where the zero-crossing slope and, therefore, the magnetic-field sensitivity is maximized.
    }
    \label{fig:sensitivity_optimization}
\end{figure}

Optimizing all eight ODMR resonances for their best sensitivity simultaneously is generally not possible.
Due to imperfect optical and MW addressing, each resonance will have a slightly different contrast and linewidth and, therefore, a different set of optimal measurement parameters, especially when one changes laser power, MW power, MW modulation frequency, or MW modulation deviation.
FIG. \ref{fig:sensitivity_optimization} shows, as an example, the dependence of the LIA signal slope on the MW modulation frequency and deviation.

This leads to the fact that each resonance has its own sensitivity.
Table \ref{tab:sensitivities} lists our achieved sensitivities for each resonance of the ODMR spectrum shown in FIG. \ref{fig:odmr_lia} and \ref{fig:odmr_spcm}.

\begin{table}[htbp]
    \caption{Magnetic-field sensitivities for each resonance}
    \centering
    \begin{tabular}{c c c c}
        \hhline{====}
        Nr. & $\eta_{\text{LIA}}$ & $\eta_{\text{SN,MM}}$ & $\eta_{\text{SN,SM,MM}}$ \\
        \hline
        1 & \SI{60.0}{\nano\tesla\per\sqrt{\hertz}} & \SI{6.14}{\nano\tesla\per\sqrt{\hertz}}  & \SI{30.0}{\nano\tesla\per\sqrt{\hertz}} \\
        2 & \SI{53.8}{\nano\tesla\per\sqrt{\hertz}} & \SI{5.90}{\nano\tesla\per\sqrt{\hertz}}  & \SI{30.5}{\nano\tesla\per\sqrt{\hertz}} \\
        3 & \SI{60.9}{\nano\tesla\per\sqrt{\hertz}} & \SI{9.89}{\nano\tesla\per\sqrt{\hertz}} & \SI{48.2}{\nano\tesla\per\sqrt{\hertz}} \\
        4 & \SI{70.4}{\nano\tesla\per\sqrt{\hertz}} & \SI{7.95}{\nano\tesla\per\sqrt{\hertz}} & \SI{29.7}{\nano\tesla\per\sqrt{\hertz}} \\
        5 & \SI{77.2}{\nano\tesla\per\sqrt{\hertz}} & \SI{9.70}{\nano\tesla\per\sqrt{\hertz}} & \SI{33.9}{\nano\tesla\per\sqrt{\hertz}} \\
        6 & \SI{51.8}{\nano\tesla\per\sqrt{\hertz}} & \SI{8.17}{\nano\tesla\per\sqrt{\hertz}} & \SI{40.9}{\nano\tesla\per\sqrt{\hertz}} \\
        7 & \SI{57.5}{\nano\tesla\per\sqrt{\hertz}} & \SI{6.34}{\nano\tesla\per\sqrt{\hertz}} & \SI{30.1}{\nano\tesla\per\sqrt{\hertz}} \\
        8 & \SI{59.2}{\nano\tesla\per\sqrt{\hertz}} & \SI{6.50}{\nano\tesla\per\sqrt{\hertz}} & \SI{27.7}{\nano\tesla\per\sqrt{\hertz}} \\
        \hhline{====}
    \end{tabular}
    \label{tab:sensitivities}
\end{table}

\section{Magnetic field calculation}\label{sec:b_field_derivation}

To calculate the magnetic field from measured resonances, we solve the NV-Hamiltonian analytically, following \cite{balasubramanian_nanoscale_2008}.
Consider the shifted Hamiltonian for one NV center \cite{rondin_magnetometry_2014} with $\text{Tr}(H) = 0$
\begin{equation}
    H = DS_z^2 + E(S_x^2 + S_y^2) + \frac{\vec{B}}{\gamma}\cdot\vec{S} - \frac{2}{3}DI,
\end{equation}
where $D = \SI{2.87}{\giga\hertz}$ is the zero-field splitting, $E \approx \SI{5}{\mega\hertz}$ is the diamond dependent stress-splitting, $\vec{B}$ is the magnetic-field vector, $S_{x,y,z}$ are the $3\times3$ Pauli matrices and $I$ is the unit matrix.
Without loss of generality, the magnetic field $\vec{B} = \gamma B(\sin\theta\hat{x} + \cos\theta\hat{z})$ can always be rotated around the NV-axis to remove the $B_y$ component, where $\theta$ is the angle between the magnetic-field vector and the NV-axis.
The roots of the characteristic polynomial are then given by
\begin{multline}
    \label{equ:char_poly}
    0 = \lambda^3 - \left(\frac{1}{3}D^2 + E^2 + B^2\right)\lambda \\
    - \frac{1}{2}DB^2\cos(2\theta) - EB^2\sin^2\theta \\
    - \frac{1}{6}DB^2 + \frac{2}{27}D^3 - \frac{2}{3}DE^2.
\end{multline}
This polynomial has three real roots $\lambda_1$, $\lambda_2$ and $\lambda_3$
\begin{multline}
    \label{equ:general_poly}
    (\lambda - \lambda_1)(\lambda - \lambda_2)(\lambda - \lambda_3) = \lambda^3 - (\lambda_1+\lambda_2+\lambda_3)\lambda^2 \\
    + (\lambda_1\lambda_2+\lambda_2\lambda_3+\lambda_1\lambda_3)\lambda - \lambda_1\lambda_2\lambda_3.
\end{multline}
When defining upper and lower resonance frequencies as $f_u = \lambda_3 - \lambda_1$ and $f_l = \lambda_2 - \lambda_1$, these roots can be expressed as $\lambda_1 = (-f_u-f_l)/3$, $\lambda_2 = (2f_l-f_u)/3$ and $\lambda_3 = (2f_u-f_l)/3$ respectively.

We compare the individual terms of Equ. \ref{equ:char_poly} and \ref{equ:general_poly} to derive formulas for $B^2$ and $\cos^2\theta$. 
Comparing linear terms leads to
\begin{equation}
    \label{equ:b_abs}
    B^2 = \frac{1}{3} \left(f_u^2 + f_l^2 - f_uf_l - D^2 - 3E^2\right),
\end{equation}
while comparing constant terms leads to
\begin{multline}
    \label{equ:b_theta}
    \cos^2\theta = \frac{2f_l^3-3f_l^2f_u-3f_lf_u^2+2f_u^3}{27(D-E/2)B^2} \\
    + \frac{2D^3 - 18DE^2}{27(D-E/2)B^2} + \frac{2D-3E}{6(D-E/2)}.
\end{multline}
After measuring both resonance frequencies $f_{u,j}$ and $f_{l,j}$ for all four NV-orientations $j$, Equ. \ref{equ:b_abs} gives us four independent measurements of the absolute value of the magnetic field.
We can simply take the mean value of all four $B_j$ as our final result $B$.

We now only know the angles $\theta_{j}$ between the magnetic-field vector and the NV-axis $j$.
This is akin to four cones in three-dimensional space, where we want to find the vector that best overlaps all four cones \cite{wang_orientation_2023}.
The magnetic field direction can be obtained by considering the scalar product $\hat{n}\cdot\hat{b} = \cos\theta$ between the NV-axis unit vector $\hat{n}$ and the magnetic-field unit vector $\hat{b}$.
This gives us a linear over-determined system $Nb = c$ of four equations with three unknowns that can be written in matrix form as
\begin{equation}
    \frac{1}{\sqrt{3}} \begin{pmatrix}
        1 & 1 & 1  \\
        1 & -1 & -1 \\
        -1 & 1 & -1 \\
        -1 & -1 & 1 \\
    \end{pmatrix} \cdot \begin{pmatrix}
        b_x \\
        b_y \\
        b_z \\
    \end{pmatrix} = \begin{pmatrix}
        \cos\theta_1 \\
        \cos\theta_2 \\
        \cos\theta_3 \\
        \cos\theta_4 \\
    \end{pmatrix},
\end{equation}
where the matrix $N$ is constituted of the unit vectors along the NV-axes, which are parallel to $[111]$, $[1\bar{1}\bar{1}]$, $[\bar{1}1\bar{1}]$ and $[\bar{1}\bar{1}1]$ respectively \cite{schloss_simultaneous_2018}.

The solution to this kind of linear least-square problem is known to be $\hat{b} = (N^TN)^{-1}N^Tc$ \cite{press_numerical_2007}, which gives us our final magnetic-field vector $\vec{B} = \gamma B\hat{b}$ as
\begin{equation}
    \vec{B} = \gamma B \frac{\sqrt{3}}{4} \begin{pmatrix}
    1 & -1 & -1 & 1 \\
    1 & -1 & 1 & -1 \\
    1 & 1 & -1 & -1 \\
    \end{pmatrix} \cdot \begin{pmatrix}
    \pm\cos\theta_1 \\
    \pm\cos\theta_2 \\
    \pm\cos\theta_3 \\
    \pm\cos\theta_4 \\
    \end{pmatrix}
\end{equation}
Because of the ambiguity in $\pm\cos\theta$, we have $2^4$ possible solutions $\hat{b}$. However, most of these solutions represent four cones that are not oriented correctly. Unwanted solutions will have a large sum of squared residuals \cite{press_numerical_2007}
\begin{equation}
    S(\hat{b}) = (c-N\hat{b})^T (c-N\hat{b}),
\end{equation}
meaning we can compute $S(\hat{b})$ and choose the vector $\hat{b}$ that minimizes $S$.

Finally, the resulting vector $\vec{B}$ can only be determined up to a symmetry of $T_d$, stemming from the tetrahedral structure of the diamond lattice.
One can circumvent this restriction with a known bias field, which breaks the symmetry of the problem.
After determining the direction of the bias field using a known external magnetic field, changes in the magnetic field that are small compared to the bias field can be vectorially resolved without ambiguities.
The resulting magnetic-field vector still has to be transferred from NV coordinates to outer lab coordinates.
This can also be accomplished with a known bias field because the unknown orientation of the diamond on the fiber tip can be obtained from one calibration measurement of the known bias field.

\section{Comparison of sensors in terms of sensing volume}\label{sec:appendix E}

\begin{figure*}[htbp]
    \centering
    \includegraphics[width=\linewidth]{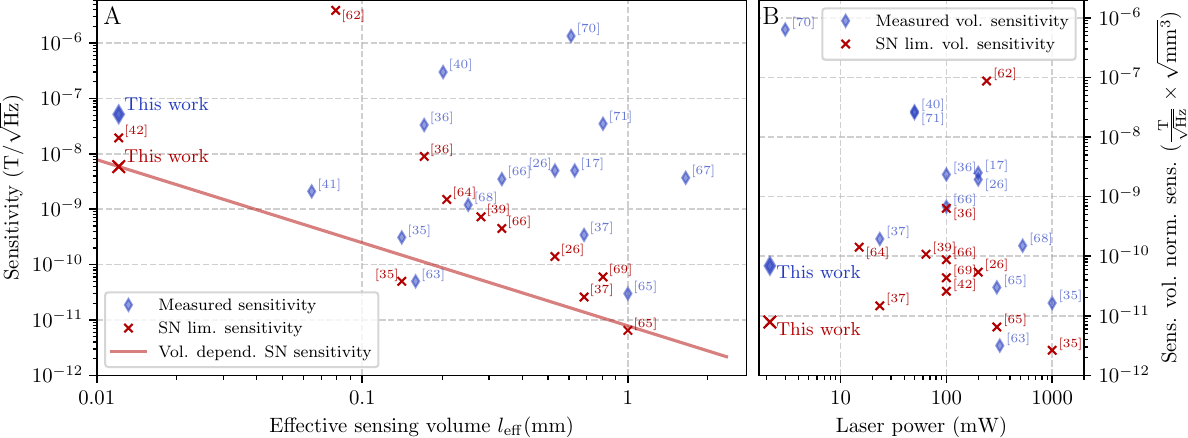}
    \caption{
        Measured and shot-noise-limited calculated sensitivities of our and other publications \cite{quan_fiber-coupled_2023, zhang_high-sensitivity_2022, dmitriev_concept_2016, graham_fiber-coupled_2023, hatano_simultaneous_2021, liu_closed-loop_2024, liu_millimeterscale_2023, blakley_room-temperature_2015, deguchi_compact_2023, blakley_fiber-optic_2016} for different effective diamond sizes with respect to the diamond or, if defined, sensing volume (A). 
        The values for measured sensitivities represent retrieved values from noise and time-trace measurements compared to shot-noise limited sensitivities, which are calculated from measured contrast and linewidth of fit data and photon count rate.
        Normalized sensitivity of the presented sensors by the individual diamond or, if given, sensing volume at the given laser power (B). 
        The bandwidth of the presented sensors is set differently in most cited publications and therefore not compared.
    }
    \label{fig:comparison_sens_vol}
\end{figure*}

In addition to FIG. \ref{fig:comparison_sens}, we plot in FIG. \ref{fig:comparison_sens_vol} the sensitivity versus the effective sensing volume, which we either retrieved directly or calculated based on the dimensions provided in the cited publications. If only the diamond volume was specified, we used this values instead as an approximation.

\FloatBarrier

\end{document}